\documentstyle[11pt,epsfig,menu99]{article}

\newcommand{\beq}{\begin{equation}}
\newcommand{\eeq}{\end{equation}}
\newcommand{\beqa}{\begin{eqnarray}}
\newcommand{\eeqa}{\end{eqnarray}}

\begin{document}
    \setlength{\baselineskip}{2.6ex}

\title{WORKING GROUP SUMMARY: ISOSPIN VIOLATION}
\author{Ulf-G. Mei\ss ner\\
{\em FZ J\"ulich, IKP (Th), D-52425 J\"ulich, Germany}}

\maketitle

\vspace{-2.5cm}

\hfill {\small FZJ-IKP(TH)-1999-28}

\vspace{1.9cm}

\begin{abstract}
\setlength{\baselineskip}{2.6ex}
\noindent I give an introduction to the problem of isospin violation
and add some comments to the various topics addressed in the
working group.

\end{abstract}

\setlength{\baselineskip}{2.6ex}

\section*{ISOSPIN VIOLATION: GENERAL ASPECTS}

\noindent Isospin symmetry was introduced in the thirties by Heisenberg
in his studies of the atomic nucleus. Since then, many particles have
been found to appear in iso--multiplets, like the nucleons, the pions,
the delta isobars, a.s.o. With the advent of QCD, a deeper understanding
of isospin symmetry has emerged. In the limit of equal up and down 
current quark
masses and in the absence of electroweak interactions, isospin is an exact
symmetry of QCD. The intra--multiplet mass splittings  allow to
quantify the breaking of this symmetry, which is caused by different
mechanisms (for a detailed review, see ref.\cite{GLrev}). 
First, the light quark masses are everything but equal (still,
their absolute masses are much smaller than any other QCD scale and thus
this breaking can be treated as a perturbation). Second, the light quarks
have different charges and thus react differently to the electromagnetic (em)
interactions. The em effects are also small since they are proportional to the
fine structure constant $\alpha = e^2/ 4\pi \approx 1/137$. In the case of
the pions, the mass splitting is almost entirely of em origin. This can be
traced back to the absence of d--like couplings in SU(2), thus promoting
the quark mass difference to a second order effect. For the nucleons,
matters are different, strong and em effects are of similar size but
different signs. The fact that the neutron is heavier than the proton
leads to the conclusion that $m_d > m_u$, consistent with the analysis of
the kaon masses.  Since we know that
isospin is broken - so why bother? First, the picture that has emerged from
the hadron masses can not be considered complete, there is still on--going
discussion about the size of the violation of Dashen's theorem, the possibility
of a vanishing up quark mass to solve the strong CP problem and ``strange''
results from lattice gauge theory. Also, the analysis of the quark
mass dependence of the baryon masses remains to be improved (for a
classic, see ref.\cite{jg} and a recent study, see ref.\cite{bm}).
Furthermore, only a few dynamical 
implications of isospin violation have been verified experimentally and a
truely quantitative picture has not yet emerged. 
In addition, the nucleus as a many--body
system offers a novel laboratory to study isospin violation.
In addition, with the advent of CW electron accelerators and improved
detectors, we now have experimental tools to measure threshold pion
photoproduction with an unprecedented accuracy.

\section*{THE PION SECTOR}

\noindent The purely mesonic sector was not touched upon in this working group,
but there is one recent result which I would like to discuss. In elastic
$\pi\pi$ scattering, the chiral perturbation analysis has been carried out
to two loops. It was demonstrated in refs.\cite{MMS,KU} that the 
em isospin--violating effects are of the same size as the hadronic two--loop
corrections. For a precise description of low energy pion reactions, it is
thus mandatory to include such effects consistently. A somewhat surprising
result was found in case of the scalar and the vector form factor of the
pion in ref.\cite{dilbert}. It was shown that the em corrections to the
momentum--dependence of both form factors are tiny (due to large
cancellations between various contributions), much smaller than the
corresponding hadronic two--loop contributions worked out in 
refs.\cite{GM,BT}. This result remains to be understood in more
detail. It is particularly
surprising for the scalar form factor since it is not protected by a
conserved current theorem \`a la Ademello--Gatto. Only the normalization
of the scalar form factor exhibits the few percent em corrections anticipated
from the study of the $\pi\pi$ scattering lengths. Note, however, that
the smallness of the effects of the light quark mass difference for
the pion form factors has been known and understood since long~\cite{GLAnn}.

\section*{THE PION--NUCLEON SECTOR}

\noindent The pion--nucleon system plays a particular role in the
study of isospin violation. First, the explicit chiral symmetry breaking and
isospin breaking operators appear at the same order in the effective
Lagrangian which maps out the symmetry breaking part of the QCD
Hamiltonian, i.e. the quark mass term (restricted here to the two lightest flavors), 
\begin{equation}
{\cal H}_{\rm QCD}^{\rm sb} = m_u \bar{u}u + m_d \bar{d}d
= \frac{1}{2}(m_u + m_d) ( \bar{u}u +  \bar{d}d ) +
\frac{1}{2}(m_u - m_d) ( \bar{u}u -  \bar{d}d )~,
\end{equation}
so that the strong isospin violation is entirely due to the isovector
term whereas the isoscalar term leads to the explicit chiral symmetry
breaking. In the presence of nucleons (and in contrast to the pion
case), both breakings appear at the
same order. This can lead to sizeable isospin violation as first
stressed in reactions involving neutral pions by
Weinberg~\cite{weinmass}. Let me perform some naive dimensional
analysis for the general case (say for any given channel in $\pi N$
scattering that is not suppressed to leading
order). Isospin--violation (IV) should be of the size
\begin{equation}\label{dim}
{\rm IV} \sim \frac{m_d-m_u}{\Lambda_{\rm hadronic}} \approx \frac{m_d
  -m_u}{M_\rho} = {\cal O}(1\%)~,
\end{equation}
where the mass of the $\rho$ set the scale for the non--Goldstone physcis.
In the presence of a close--by and strongly coupled baryonic resonance like the
$\Delta (1232)$, IV  might be enhanced
\begin{equation}
{\rm IV} \sim \frac{m_d-m_u}{m_\Delta - m_N} = {\cal O}(2\%)~.
\end{equation}
Of course, such type of arguments can not substitute for full scale
calculations. Second, there are two analyses~\cite{glk,mats}
which seem to indicate a fair amount of isospin violation (of the
order of 6...7\%, which is much bigger than the dimensional arguments
given above would indicate) in
low--energy $\pi N$ scattering, see Gibbs' talk~\cite{billZ}. This
can not be explained in conventional meson--exchange models by standard
meson mixing mechanisms. I would also like to mention that in these
two analyses the hadronic and the electromagnetic contributions are
derived from different models. This might cause some concern about
possible uncertainties due to a theoretical mismatch. Clearly, it
would be preferable to use here one unique framework. That can, in
principle, be supplied by chiral perturbation theory since
electromagnetic corrections can be included systematically by a
straightforward extension of the power counting. This is most economically,
done by counting the electric charge as a small
parameter, i.e. on the same footing as the external momenta and meson masses.
The heavy baryon chiral perturbation theory machinery to study these
questions to  complete one--loop (fourth) order has been set up as shown by
M\"uller~\cite{guidoZ}. It is important to perform such calculations
to fourth order since one--loop graphs appear at dimension three {\it
  and} four. Furthermore, it is known from many studies that one--loop
diagrams with exactly one insertion from the dimension two $\pi N$
Lagrangian are (often) important. Finally, symmetry breaking (chiral
and isospin) in the loops only starts at fourth order.
In particular, questions surrounding the $\pi
N$ $\sigma$--term or neutral pion scattering off nucleons can now be
addressed to sufficient theoretical precision. A first step in this
direction for all channels in $\pi N$ scattering was reported by
Fettes~\cite{nadiaZ}, but a full scale one--loop calculation
including all virtual photon effects still has to be done. Of
particular interest is the novel relation between $\pi^0$ and
$\pi^\pm$ scattering off protons that is extremely sensitive to
isospin violation. It should also be stressed that for such tests, it
is mandatory to better measure and determine the small isoscalar $\pi
N$ amplitudes. Also, the relations which include the much bigger
isovector amplitudes show IV consistent with the dimensional arguments
given in eq.(\ref{dim}). I
consider the ``ordering schemes'' discussed by Gibbs and Fettes very
useful tools to pin down the strengths and sources of isospin breaking
in $\pi N$ scattering. This also allows to see a priori which type of
measurements are necessary to obtain complete information and to what
extent various reactions can give redundant information (one example
is discussed by Gibbs~\cite{billZ}). Intimately related to this is
pion--photoproduction via the final--state
theorem, i.e. certain $\pi N$ scattering phases appear in the
imaginary part of the respective charged or neutral pion photoproduction multipoles.
Bernstein~\cite{aronZ} stressed that in neutral pion
photoproduction off protons, there are two places to look for isospin
violation. One is below the $\pi^+ n$ threshold, which might give
access to the elusive (but important) $\pi^0 p$ scattering length. At
present, it does not appear that the original proposal of measuring
the target polarization below the $\pi^+ n$ threshold to high
precision is feasible at a machine like e.g. MAMI. The
other important effect, which appears to be more easily accessible to an
experiment, is the strength of the cusp at the opening of the $\pi^+ n$
threshold, which according to Bernstein's three--channlel S--matrix analysis~\cite{aronZ} is
quite sensitive to isospin violation. Such a calculation should also be done
in the framework of heavy baryon chiral perturbation theory (beyond
the charged to neutral pion mass difference effects included so far). 
Over the last years, there has been a very fruitful interplay between
experimenters and theorists particularly in the field of pion photo--
and electroproduction and it is of utmost important to further
strengthen this. It is a theorists dream
that  reactions with neutral pions (elastic scattering and
photoproduction) will be measured to a high precision.
An important point was stressed by Lewis~\cite{randyZ}. In a ``toy''
calculation (i.e. an SU(2) approach to the strange vector form factor
of the nucleon, which is clearly related to three flavor QCD),
 he showed that isospin--breaking effects can simulate
a ``strange'' form factor that intrinsically vanishes in that approach.
This nicely demonstrates that to reliably determine small quantities, may they
be related to isospin conserving or violating operators, {\it all}
possible effects have to be included. The recent measurements at BATES
and JLAB, which seem to indicate small expectation values of the
strange vector current in the proton, should therefore be reanalyzed. In this
case, isospin violation appears to be a nuisance but can not be ignored.

\section*{THE NUCLEON--NUCLEON SECTOR}

\noindent The only new data with respect to IV were presented by
Machner~\cite{machZ}. He analysed recent data from COSY and IUCF
for $pp \to \pi^+ d$ and $np \to \pi^0 d$. For exact isospin symmetry
(i.e. after removing the Coulomb corrections), the pertinent cross
sections should be equal (up to a Clebsch). In the threshold region,
one can make a partial wave expansion and finds that the S--wave
contribution $\alpha_0$ shows IV of the order of 10\% and no effect
is observed in the P--wave terms. To my knowledge, a theoretical
understanding of this effect is lacking. Despite a huge amount of
efforts over the last years, a model--independent effective field
theory description of pion production in proton--proton collisions
has not yet been obtained. The energies involved to even produce a
pion at rest are too large for the methods employed so far.
More progress, however, has been made in the two--nucleon system at low energies. 
It is well known that IV appears in the NN scattering lengths. In the
nuclear jargon, one talks about charge independence breaking (CIB) ($a_{np}
\neq (a_{pp} + a_{nn})/2$ after Coulomb subtraction, where $a$ denotes
the scattering length) and charge symmetry breaking (CSB) ($a_{pp} \neq
a_{nn}$ after Coulomb subtraction). These effects are naturally most
pronounced at threshold.
Kaplan, Savage and Wise (KSW)~\cite{ksw} have proposed a
non--perturbative scheme that allows for power counting on the level
of the nucleon--nucleon scattering amplitude. In that framework, IV
(CIB and CSB) has recently been investigated~\cite{em}.
It was shown that isospin violation can be systematically included
in the effective field theory approach to the two--nucleon system in
the KSW formulation. For that, one has to construct the most general
effective Lagrangian containing virtual photons and extend the
power counting accordingly.
This framework allows one to systematically classify the various
contributions to CIB and CSB. In
particular, the power counting combined with dimensional analysis 
allows one to understand the suppression of contributions
from a possible charge--dependence in the pion--nucleon coupling constants.
Including the pions, the leading CIB breaking effects are the
pion mass difference in one--pion exchange together with a four--nucleon contact term.
These effects scale as $\alpha {Q}^{-2}$, where $Q \approx 1/3$ is the
genuine expansion parameter of the KSW scheme. Power counting lets one
expect that the much debated contributions from two--pion exchange and
$\pi\gamma$ graphs are suppressed by factors of $1/3$ and $(1/3)^2$,
respectively. This is in agreement with some, but not all, previous
more model--dependent calculations. The leading charge
symmetry breaking is simply given by a four--nucleon contact term. 

\section*{LIGHT NUCLEI}

\noindent Often, the nucleus can be used as a filter to enhance or
suppress certain features of reactions as they appear in free space. Furthermore,
measurements on the neutron, which are  necessary to get the
complete information in the isospin basis (for a discussion on this
topic with respect to pion photoproduction, see ref.\cite{ulfo}) can
only be done on (preferably polarized) light
nuclei. Gibbs~\cite{billZ} has pointed out that a measurement of
charge exchange on the proton and the neutron (in forward direction
and close to the interference minimum near 45~MeV) could be done in the
$^3$He--triton system. This would be an interesting possibility to get
another handle on the elusive neutron and allow one to pin down one
of the amplitudes parametrizing IV (according to the ordering scheme
mentioned above). For a more detailed discussion
concerning the extraction of neutron properties from the deuteron, I refer to the
recent summary by Beane~\cite{silas}.

\section*{WHERE DO WE STAND AND WHERE TO GO}

\noindent For sure, isospin symmetry is broken. However, do we
precisely know the size of IV from experiment? The answer is yes and
no. We have some indicative information but no systematic
investigations of all pertinent low energy reactions are available.
Also, one might ask the question whether the methodology, which has
been used so far to extract numbers on IV, say from low energy $\pi N$
scattering data, is reliable?  If we assume that this is the case, we still
have no deeper understanding of the mechanisms triggering IV. To my
knowledge, the only  machinery to consistently separate
strong and electromagnetic IV is based on effective field theory. In
that scheme, one can consider various reactions like elastic $\pi N$
scattering, pion photoproduction or even nucleon Compton scattering to
try to get a handle on the symmetry breaking operators $\sim m_d
-m_u$. Also, a systematic treatment of isospin violation is mandatory for
the determiantion of small quantities like the isoscalar S--wave
scattering length or the strange nucleon form factors. Based on that,
I have the following wish list for theory and experiment:

\medskip

\noindent {THEORY}:

\begin{enumerate}
\item[$\bullet$] The effective chiral Lagrangian calculations can and
need to be improved. In particular, it is most urgent to get a handle
on the so--called low--energy constants, which parametrize the
effective Lagrangian beyond leading order. Sum rules, models or even
the lattice might be useful here.
\item[$\bullet$] A deeper theoretical understanding of certain
phenomenological models (like e.g. the extended tree level model
of ref.\cite{ET}) in connection with the approaches to correct for
Coulomb effects would be helpful.
\item[$\bullet$] The dispersion--theoretical approach should be
  revisited and set up in a way to properly include IV (beyond what
  has been done so far). For some first steps, see the talk by 
  Oades~\cite{oadesZ}.
\end{enumerate}  

\smallskip

\noindent {EXPERIMENT}:
\begin{enumerate}
\item[$\bullet$] Clearly, we need more high precision data for the
  elementary processes, but not only for
  $\pi N$ scattering but also for (neutral) pion photo/electroproduction.
\item[$\bullet$] More precise nuclear data are also needed. Embedding
  the elementary reactions in the nucleus as a filter allows one to get
  information on the elusive neutron properties. Clearly, this refers
  to few--nucleon systems where precise theoretical calculations are
  possible.
\end{enumerate} 

\noindent Finally, I would like to stress again that a truely
  quantitative understanding of isospin violation can only be obtained
  by considering a huge variety of processes. While pion--nucleon
  scattering is at the heart of these investigations, threshold pion
  photoproduction or the nucleon form factors also play a vital role
  in supplying additional information. 

\section*{ACKNOWLEDGEMENTS}

\noindent I would like to thank all participants of this working group
for their contributions. I am grateful to my collaborators Evgeny
Epelbaum, Nadia Fettes, Bastian Kubis, Guido M\"uller and Sven
Steininger for sharing with me their insight into this topic.
Last but not least the superbe organization by Christine Kunz and 
Res Badertscher is warmly acknowledged.

\bibliographystyle{unsrt}

\end{document}